\documentclass[twocolumn,aps,prb,reprint,longbibliography,floatfix]{revtex4-2} 

\usepackage[T1]{fontenc}
\usepackage{amsmath,amssymb,mathtools}
\usepackage{graphicx}
\usepackage{bm}

\usepackage{hyperref}
\hypersetup{%
	pdftitle={Diffusive and hydrodynamic magnetotransport around a density perturbation
		in a two-dimensional electron gas},%
	pdfauthor={P. Shubham Parashar and Michael M. Fogler},%
	pdfpagemode={UseNone},%
	pdfstartview={FitH},%
	breaklinks=true,%
	citecolor=blue,%
	colorlinks=true,%
	linkcolor=blue,%
	urlcolor=blue
}

\newcommand{\vect}[1]{\mathbf{#1}}

\newcommand{\Rdip}{R_L}
\newcommand{\thetadip}{\theta_L}
\newcommand{\RnogoO}{R_\mathrm{no}} 
\newcommand{\RnogoH}{R_\mathrm{no}}
\begin{document}
\title{Diffusive and hydrodynamic magnetotransport around a density perturbation
in a two-dimensional electron gas}

\author{P. Shubham Parashar}
\author{Michael M. Fogler}
\affiliation{Department of Physics, University of California San Diego, La Jolla, California 92093, USA}

\date{\today}

\begin{abstract}

We study current flow around a density inhomogeneity in a two-dimensional electron gas in the presence of a strong magnetic field.
The inhomogeneity is parametrized by a power-law tail with an exponent $\beta > 2$.
We show that current and electrochemical potential are exponentially suppressed inside a surrounding area much larger than the geometric size of the perturbation.
The corresponding ``no-go'' radius grows as a certain power of the magnetic field.
Residual current and potential exhibit spiraling patterns inside the no-go region.
Outside of it, they acquire corrections inversely proportional to the distance,
which is known as the Landauer resistivity dipole.
The Landauer dipole is rotated by the angle $\pi (1 - 1 / \beta)$ with respect to the average electric field.
The rotation direction depends on whether the local density is raised or lowered.
We also consider the effect of electron viscosity and show that the variation of the no-go radius with magnetic field becomes more rapid if viscosity is large enough.
The Landauer dipole size is set by the Gurzhi length,
which is much larger than the no-go radius, which is in turn much larger than the geometric size of the perturbation.
Our results may be useful for interpreting nanoimaging of current distribution in graphene and other two-dimensional systems.

\end{abstract}

\maketitle

%=========================================================
\section{Introduction}
\label{sec:intro}
%=========================================================

\begin{figure}[b]
	\centering
	\includegraphics[width=2.9in]{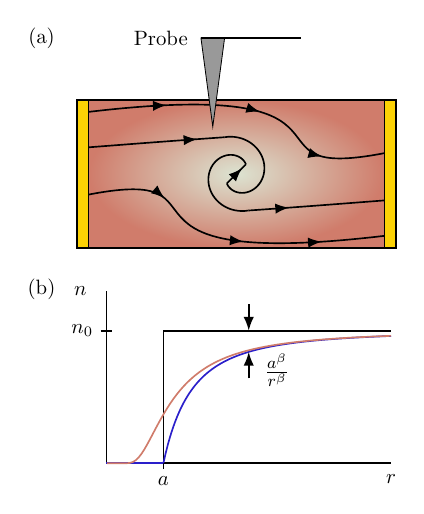}
	\caption{(a) Schematic of the experiment that motivated this study. A local probe measures the electrochemical potential produced by a current flowing around a density depletion in a 2D electron gas. In the presence of a magnetic field,
		the current forms a spiral pattern around the obstacle.
		(b) Examples of density profiles $n(r)$ studied: abrupt (black line), smooth with a flat insulating core (blue), and smooth everywhere (red line).
	}
	\label{fig:spirals1}
\end{figure}

Direct imaging of electron transport in two-dimensional (2D) systems
has become possible recently with the development of local probes capable of nanometer-scale resolution. 
Scanning single-electron transistors can measure local compressibility and potential~\cite{Yacoby1999, Martin2007, Ilani2004, Kumar2022}.
Subsurface charge accumulation imaging and scanning capacitance microscopy have been used to image local potentials and dissipation~\cite{Tessmer1998, Finkelstein2000, Suddards2012}.
Scanning atomic force microscopy~\cite{McCormick1999}, scanning tunneling microscopy and potentiometry~\cite{Muralt1986, Hashimoto2008, Li2013},
scanning SQUID~\cite{Uri2019, Uri2020}, and nitrogen-vacancy center magnetometry~\cite{Sulpizio2019, Palm2024}
are also among the techniques that provide access to the distributions of the electrochemical potential $e\Phi$ and current density $\vect{j}$.
Strong magnetic field has been shown to have a dramatic effect on electron transport.
This regime can be defined by the condition that the Hall conductivity $\sigma_{xy}$ exceeds
the dissipative conductivity $\sigma_{xx}$.
In particular, in the quantum Hall effect (QHE) the latter vanishes,
and as a result, current can flow only along the edges.
In real QHE devices, the longitudinal conductivity $\sigma_{xx}$ is small but finite, so the current penetrates into the bulk~\cite{Rendell1981, Macdonald1983}.
The current distribution is nontrivial and highly sensitive to device geometry and 
inhomogeneities which can be modeled by spatially varying $\sigma_{xx}$ and $\sigma_{xy}$~\cite{Ruzin1993}.

In this work, we study the current flow around a local density perturbation. It is motivated by recent scanning tunneling potentiometry (STP) experiments in graphene~\cite{Willke2017, Behn2021, Krebs2023, Krebs2026}.
In those experiments, a local charging effect was employed to create a circular depletion region
containing a $p$--$n$ junction of radius $0.5\text{--}1.0\,\mu\mathrm{m}$.
The depletion played a role of a strong obstacle for the current.
The maps of the electrochemical potential $e\Phi(\vect{r})$, $\vect{r} = (x, y)$ near the junction were measured
and the principal observations were as follows.
Inside the depletion, $\Phi(\vect{r})$ had not much variation indicating that the current was blocked.
The potential outside the $p$--$n$ junction was fitted to the form
\begin{equation}
\Phi(r, \theta) \propto -r \cos \theta
- \frac{\Rdip^2}{r} \cos\left(\theta + \thetadip\right) ,
\label{eqn:Phi_far}
\end{equation}
where the first term corresponds to a uniform electric field in the $\hat{\mathbf{x}}$-direction and
the second is a correction known as the Landauer resistivity dipole~\cite{Landauer1957, Landauer1975, Landauer1978, Sorbello1981, Zwerger1991, Reuss1996, Briner1996}.
(A 2D dipole potential behaves as $1 / r$ rather than $1 / r^2$ in three dimensions.
Here $r = |\vect{r}|$ is the radial distance and $\theta$ is the polar angle.)
In the absence of the magnetic field, $B = 0$, the dipole had the angular orientation $\thetadip = 0$,
consistent with a simple qualitative picture where the current incident on the obstacle causes
charge crowding (uncrowding) on its upstream (downstream) sides.
In the presence of a weak magnetic field, $B \lesssim 1\,\mathrm{T}$, the dipole rotated with respect to the asymptotic direction of the electric field: $\thetadip \neq 0$.
This implies a twisted, spiral-like current flow around the depletion,
Fig.~\ref{fig:spirals1}(a).
Additionally, the region where the current was suppressed grew beyond the geometric boundary of the $p$--$n$ junction and acquired some spiral-like structure.
These observations were analyzed using several theoretical approaches: an analytical model of diffusive transport with the hard-wall boundary condition at the $p$--$n$ junction,
a semiclassical ballistic correction to this model, and numerical simulations of quantum transport.
The first one predicted that the Landauer dipole rotation angle is $\thetadip = 2\theta_H$
where $\theta_H = \arctan(\sigma_{xy}/\sigma_{xx})$ is the Hall angle,
in agreement with the experiment (see also Refs.~\cite{Landauer1978, Reuss1996}).
The last two approaches explained an additional feature of $\Phi(x,y)$ detected near the obstacle in strong fields $B \gtrsim 1\,\mathrm{T}$: a small kink at the distance of one cyclotron diameter away from the $p$--$n$ boundary.

The hard-wall boundary condition corresponds approximately to a step-like electron density profile $n(r)$, e.g., density jumping from zero to a nonzero value at some distance $r = a$ from the center of the depletion [the black curve in Fig.~\ref{fig:spirals1}(b)].
In this work, we improve this approximation by considering a more realistic and more general model where $n(r)$ approaches its asymptotic value $n_0$ gradually [the blue and red curves in Fig.~\ref{fig:spirals1}(b)], so that the correction $n(r) - n_0$ has a power-law tail:
\begin{equation}
\frac{n(r)}{n_0} \simeq
1 - \left(\frac{a}{r}\right)^{\beta}, \quad r\gg a .
\label{eqn:denprof}
\end{equation}
If the density depletion is induced by local doping, then
according to electrostatics~\cite{Fogler2007}, $\beta = 3$ at large $r$.
Other values of $\beta$ may be possible to realize using electrostatic doping
with gate dielectrics of varying thickness or electrodes of special shape.
We will assume that the power-law decrease is fast enough, $\beta > 2$.
Under this condition, the depletion acts as a local perturbation
so that the asymptotic dipolar law [Eq.~\eqref{eqn:Phi_far}] is valid.

Our analysis shows that the size $\Rdip$ and the rotation angle $\thetadip$ of the Landauer resistivity dipole are qualitatively different
in the step-like and the gradual density profile cases.
In the latter, $\Rdip$ is much larger than $a$ and grows with $B$ if the magnetic field is strong.
This implies that even a weak density depletion can strongly repel the injected current.
The size of the ``no-go'' region, i.e., the region where the current is strongly suppressed compared to its asymptotic value at infinity is of the order of $\Rdip$ rather than $a$ and it grows with $B$.
Remarkably, the Landauer dipole rotation angle is reduced compared to that for the hard-wall case: $\thetadip \simeq 2\theta_H - (\pi / \beta)$.
We also briefly examine the case of density accumulation, where the sign of the correction
in Eq.~\eqref{eqn:denprof} is positive, in which case the Landauer dipole rotates in the opposite direction.

It has been shown theoretically~\cite{Lucas2017} that the size of the Landauer dipole is determined primarily by the so-called Gurzhi length~\cite{Gurzhi1968} if the electron fluid has high viscosity.
This effect is related to the Stokes paradox in the context of a hydrodynamic flow around a cylinder~\cite{Lamb2005, Landau1987}.
Hydrodynamic flow around a hard obstacle in a 2D electron gas in the presence of a magnetic field has been studied in Refs.~\cite{Gornyi2023, Alekseev2023}.
In particular, the charge density distribution of the induced Landauer dipole was investigated in Ref.~\cite{Gornyi2023}.
High-viscosity electron fluid can be realized in modern high-mobility 2D systems,
such as semiconductor nanostructures~\cite{Molenkamp1994, Jong1995, Levin2018, Gusev2018, Gusev2020, Gupta2021, Keser2021, Wang2022} and graphene~\cite{Bandurin2016, Crossno2016, KrishnaKumar2017, Bandurin2018, Berdyugin2019, Sulpizio2019, Ku2020, Geurs2020},
and also in some quasi-2D metals~\cite{Moll2016, Bachmann2022, Gooth2018, Vool2021, AharonSteinberg2022}
where electron-electron scattering rate is much higher than electron-impurity one.
Motivated by these developments, we examine the role of viscosity in our magnetotransport problem where the electron density is nonuniform, Eq.~\eqref{eqn:denprof}.
We show that in the viscosity-dominated regime,
the scaling of the no-go radius as a function of $B$ is modified compared to the diffusion-dominated one: it is characterized by a significantly larger power-law exponent.
The size of the Landauer dipole is given by the Gurzhi length,
with a logarithmic correction, similar to the zero-field case.

The remainder of this paper is organized as follows.
In Sec.~\ref{sec:diffusive} we study the diffusive regime.
In Sec.~\ref{sec:hydro} we analyze the viscous flow regime.
We conclude with a brief discussion in Sec.~\ref{sec:discussion}.
Details of the derivations are presented in the Appendix.

%=========================================================
\section{Diffusive transport}
\label{sec:diffusive}
%=========================================================

\subsection{Model}

In the diffusive regime the current density obeys the Ohm's law,
\begin{equation}
\vect{j}(\vect{r}) = \sigma_{xx}(\vect{r}) \vect{E}(\vect{r})
- \sigma_{xy}(\vect{r}) \left[\hat{\vect{z}} \times \vect{E}(\vect{r})\right]
\,.
\label{eqn:j_from_E}
\end{equation}
It is common to refer to $\vect{E}(\vect{r})$ as the electric field.
However, strictly speaking, $\vect{E}(\vect{r}) = -\bm{\nabla} \Phi(\vect{r})$ where
$e\Phi(\vect{r}) = e\Phi_e(\vect{r}) + \varepsilon_F(\vect{r})$ is the electrochemical potential,
which is the sum of the electrostatic energy per electron $e\Phi_e(\vect{r})$ and the
local chemical potential $\varepsilon_F(\vect{r})$.
The usual transport measurements and most of the scanned probe
experiments measure $\Phi(\vect{r})$ rather than $\Phi_e(\vect{r})$.

To make the model simple, we use the Drude model for the conductivity:
\begin{align}
\sigma_{xy}(\vect{r}) &= \frac{e^2 \tau_\mathrm{mr}}{m} \frac{\alpha}{1 + \alpha^2}\, n(\vect{r}),
\label{eqn:sigmaxx-def}\\
\sigma_{xx}(\vect{r}) &= \alpha\, |\sigma_{xy}(\vect{r})|\,,
\\
\alpha &= \frac{1}{\omega_{c} \tau_\mathrm{mr}}\,,
\label{sigmadef}
\end{align}
where $m$ is the electron effective mass and
$\omega_{c} = e B / (m c)$ is the cyclotron frequency.
In a 2D electron system with a strongly nonparabolic band dispersion, such as graphene,
$m$ is a function of $\varepsilon_F = \varepsilon_F(n)$,
so strictly speaking the Drude model can be used only if the relevant range of
$n$ is narrow. In the case of graphene,
this means that the background electron density $n_0$ should not be too low.
Parameter $\tau_\mathrm{mr}$ in Eq.~\eqref{sigmadef} is the momentum-relaxation time.
Typically, $\tau_\mathrm{mr}$ decreases with the magnetic field
as the QHE is approached.
Assuming a simplified power-law dependence $\tau_\mathrm{mr} \propto B^{-\chi}$ with some $0 < \chi < 1$
as a first-order approximation, we find then $\alpha \propto 1/B^{1 - \chi}$.
The inverse of Eq.~\eqref{eqn:j_from_E} is
\begin{equation}
\vect{E}(\vect{r}) = \rho_{xx}(\vect{r})\, \vect{j}(\vect{r})
- \rho_{xy}(\vect{r}) \left[\hat{\vect{z}} \times \vect{j}(\vect{r})\right]
\,,
\label{eqn:E_from_j}
\end{equation}
where
\begin{equation}
\rho_{xy} = -\rho_{yx}
= -\frac{1}{\alpha}\, \frac{m}{e^2 n \tau_\mathrm{mr}}\,,
\quad
\rho_{xx} = \alpha |\rho_{xy}|\,.
\label{eqn:rho_xx}
\end{equation}
The current conservation condition $\bm{\nabla} \cdot \vect{j}(\vect{r}) = 0$
yields the equation for $\Phi(\vect{r})$:
\begin{equation}
\bm{\nabla} \cdot \left[\sigma_{xx}(\vect{r}) \bm{\nabla} \Phi(\vect{r})\right]
+ \left[\hat{\vect{z}} \times \bm{\nabla}\sigma_{xy}(\vect{r})\right]
\cdot \bm{\nabla} \Phi(\vect{r}) = 0\,.
\label{eqn:ohmslawcomponents}
\end{equation}
A complementary equation can be written for the
stream function $\Psi(\vect{r})$ defined by
\begin{equation}
  \vect{j}(\vect{r}) = \hat{\vect{z}} \times \bm{\nabla}\Psi(\vect{r}) .
  \label{eqn:stream-def}
\end{equation}
Using $\bm{\nabla} \times \vect{E} = 0$, we obtain
\begin{equation}
\bm{\nabla} \cdot \left[\rho_{xx}(\vect{r}) \bm{\nabla} \Psi(\vect{r})\right]
+ \left[\hat{\vect{z}} \times \bm{\nabla}\rho_{xy}(\vect{r})\right]
\cdot \bm{\nabla} \Psi(\vect{r}) = 0\,,
\label{eqn:stream_equation}
\end{equation}
which has the same structure as Eq.~\eqref{eqn:ohmslawcomponents}.
In this Section we choose to work with Eq.~\eqref{eqn:ohmslawcomponents} because it is perhaps more familiar than Eq.~\eqref{eqn:stream_equation} but in Sec.~\ref{sec:hydro} we switch to the stream-function formalism.

We are mainly interested in the high-field regime $\omega_c \tau_\mathrm{mr} \gg 1$, so that $\alpha \ll 1$ and $\sigma_{xx} \ll |\sigma_{xy}|$.
Below we discuss exact solutions of Eq.~\eqref{eqn:ohmslawcomponents} that can be found for some $n(r)$ profiles.
One such example is Eq.~\eqref{eqn:denprof} or more precisely,
\begin{equation}
\frac{n(r)}{n_0} = 1 - \left(\frac{a}{r}\right)^{\beta}, \quad r > a
\label{eqn:denprof2}
\end{equation}
supplemented with $n(r) = 0$ at $r < a$,
i.e., a flat insulating core [the blue curve in Fig.~\ref{fig:spirals1}(b)].
Another example is
\begin{equation}
n(r) = n_0 \exp \left[ -\left({a}/{r}\right)^{\beta} \right] ,
\label{eqn:denprof1}
\end{equation}
which has the same large-distance behavior but is smooth everywhere.
Both models will be shown to give the same results for the current and potential distributions in the region of primary interest $r \gg a$.

\subsection{Qualitative discussion}
\label{sub:advection}

In strong magnetic fields the current tends to flow along the lines of constant $\sigma_{xy}(\vect{r})$. One way to understand this ``Ruzin's principle''~\cite{Ruzin1993} is to invoke the analogy to advection-diffusion of a passive tracer in a hydrodynamic flow.
Indeed, Eq.~\eqref{eqn:ohmslawcomponents} is mathematically identical to the equation that governs the steady state of a system with the diffusion constant $D = \sigma_{xx}$ and the flow velocity $\vect{v} = \hat{\vect{z}} \times \bm{\nabla}\sigma_{xy}(\vect{r})$.
The contours of constant $\sigma_{xy}(\vect{r})$ are the streamlines of this flow.
The QHE regime $\sigma_{xx} \to 0$ corresponds to the limit of vanishing diffusion, in which the current is bound to the contours $\sigma_{xy}(\vect{r}) = \mathrm{const}$. In the present case, these are concentric circles around the origin.
Since these contours are closed, the transport in the bulk is impossible in the QHE regime. The current can enter or exit the system only along the distant edges where $\sigma_{xy}(\vect{r})$ eventually drops to zero, so that $\bm{\nabla}\sigma_{xy} \neq 0$.

The strong-field regime with small but nonzero $\sigma_{xx} \ll \sigma_{xy}$ maps to the case where some diffusion does occur, so that the tracer particles
can escape from the closed orbits.
This diffusive spreading is the strongest effect far away from the depletion where the velocity $\vect{v}$ is small,
so that it takes a long time for the tracer to drift any given length.
In contrast, near the depletion, where velocity $\vect{v}$ is relatively high,
the tracer tends to stay on the circular orbit for many revolutions before the diffusion moves it away.
In the context of magnetotransport, this suggests a picture where the lines of current are asymptotically straight at infinity
where $\vect{j}(x,y) = \vect{j}_0 = \mathrm{const}$
but become twisted and spiral-like~\cite{Fogler1995} near the depletion.
It also implies that there is a characteristic no-go distance $\RnogoO$
such that at $r \ll \RnogoO$ the current is strongly suppressed, $|\vect{j}| \ll |\vect{j}_0|$.
This no-go distance increases as $\sigma_{xx}$ decreases, becoming infinite in the QHE regime.

We use the following heuristic argument to estimate $\RnogoO$.
Consider a circular streamline of some radius $r$.
After one full revolution, the diffusion broadens the stream laterally by the distance
\begin{equation}
\delta(r) \sim \sqrt{2 D t}\,,
\label{eqn:difflaw}
\end{equation}
where $t = 2\pi r / |\vect{v}|$ is the travel time.
If this broadening is much smaller than $r$, then the diffusion is ineffective, i.e., the tracer remains ``locked'' on the closed orbit.
Therefore, we estimate the no-go radius $\RnogoO$ by requiring that the diffusive spreading is of the order of $\RnogoO$ itself:
\begin{equation}
  \delta(\RnogoO) \sim
  \sqrt{
  \frac{4\pi \alpha a}{\beta} 
  \left(\frac{\RnogoO}{a}\right)^{\beta + 1} \RnogoH
  } \sim \RnogoO \,.
\label{eqn:spreading1}
\end{equation}
The solution of this equation scales as
\begin{equation}
\RnogoO \sim \alpha^{-\frac1\beta} a\,.
\label{eqn:R_no1}
\end{equation}
It indicates that $\RnogoO$ is much larger than the geometric size $a$ of the depletion region provided $\alpha \ll 1$.
If $\alpha \propto 1/B^{1 - \chi} $, then the no-go radius grows with magnetic field as a weak power-law: $\RnogoO \propto B^{\frac{1 - \chi}\beta}$.

Note that the advection-diffusion analogy is a fairly faithful representation of the actual physics of the problem. The fictitious flow velocity $\vect{v}$ is proportional to the drift velocity $c \left(\vect{E}_0 \times \hat{\vect{z}}\right) / B$ of the guiding center of an electron's cyclotron orbit in crossed electric and magnetic fields.
Here $\vect{E}_0 = \bm{\nabla} \varepsilon_F / e$ is the electric field that exists in the depletion region in equilibrium.

\subsection{Analytical solutions}
\label{sub:analytical1}

We verify the above heuristic argument by solving Eq.~\eqref{eqn:ohmslawcomponents} in a regular manner.
Due to the rotational symmetry of the problem, the solution can be expanded in angular momentum harmonics $e^{i m \theta}$.
Let us assume that the electric field and current density become uniform at large $r$.
In this case, $\Phi(\vect{r})$ contains only the $m = 1$ harmonic ($p$-wave),
i.e.,
\begin{equation}
 \Phi(r,\theta) = \Re \mathrm{e}\, \left[\phi(r) e^{i \theta} \right] .
\label{eqn:potAnsatz}
\end{equation}
At large $r$, function $\phi(r)$ must have the asymptotic behavior
\begin{equation}
\phi(r) \simeq -r - \frac{\lambda}{r}\,.
\label{eqn:f_far}
\end{equation}
The first term represents a uniform unit electric field.
The second term, proportional to a complex number $\lambda$, describes the Landauer resistivity dipole with the size and angular orientation given by
\begin{equation}
\Rdip = |\lambda|^{1/2},
\quad
\thetadip = \arg \lambda\,,
\label{eqn:R_dip}
\end{equation}
cf.~Eq.~\eqref{eqn:Phi_far}.
Substituting Eq.~\eqref{eqn:potAnsatz} into Eq.~\eqref{eqn:ohmslawcomponents},
we find
\begin{equation}
\phi'' + \left(\frac{1}{r} + \frac{\sigma'_{xx}}{\sigma_{xx}}\right) \phi'
 - \left(\frac{1}{r^2} - \frac{i}{r}\frac{\sigma'_{xy}}{\sigma_{xx}} \right) \phi = 0 .
\label{eqn:radial}
\end{equation}

In addition to Eq.~\eqref{eqn:f_far}, we need a boundary condition for short distances.
For example, in the model where the conductivity vanishes at a single point [Eq.~\eqref{eqn:denprof1}],
the suitable boundary condition is $\phi(0) = 0$.
In the model with a finite-size insulating core, i.e.,
$n(r) = \sigma_{xx}(r) = \sigma_{xy}(r) = 0$ at $r \leq a$,
the potential $\phi(a)$ does not have to vanish but must be bounded.
(In fact, it is exponentially small if $\alpha \ll 1$, see below.)
To fix $\phi(r)$ at $r < a$ one needs to make additional assumptions
because Eq.~\eqref{eqn:ohmslawcomponents} becomes an identity at such $r$.
For example, if electrons are strictly forbidden to enter the core, as in the antidot problem, then $\phi(r)$ would exhibit an inverse square-root divergence
(rounded on the scale of the screening length) as $r$ approaches $a$ from inside~\cite{Fogler2004}.
To avoid such complications, we assume that the region $r < a$ has a
very small uniform conductivity [see Eq.~\eqref{eqn:denprof_abrupt} below].
In this case the electric field at $r < a$ is also uniform and the potential is linear,
\begin{equation}
\phi(r) = \frac{r}{a}\, \phi(a), \quad r < a\,.
\label{eqn:f_near_0}
\end{equation}
%%

% - - - %
\begin{figure}[t]
\centering
\includegraphics[width = 3.0in]{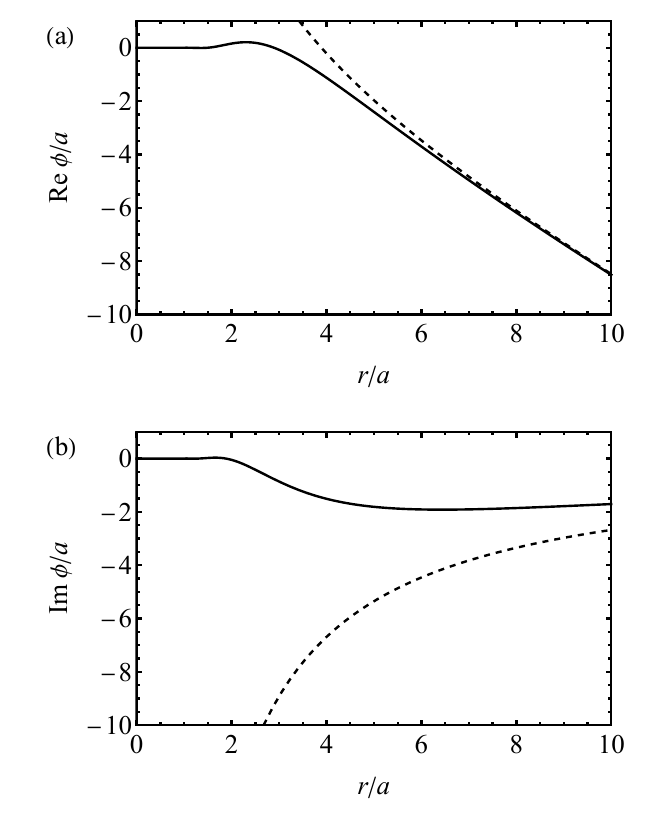}
\caption{(a) $\Re \mathrm{e}\,\phi(r)$ for $\alpha = 0.01$, $\beta = 3$. The solid line
($r/a \ge 1$ part) is Eq.~\eqref{eqn:phi_outer}, the dashed line is Eq.~\eqref{eqn:f_far}.
(b) Same for $\Im \mathrm{m}\, \phi(r)$.
\label{fig:wkb-exact-dipole}
}
\end{figure}

The exact solution of our boundary-value problem for the density profile given by Eq.~\eqref{eqn:denprof2}
can be expressed in terms of the Gauss hypergeometric functions, see
Eqs.~\eqref{eqn:fg_hypergeom}--\eqref{eqn:c_pm} in Appendix~\ref{app:beta3}.
The plots of the real and imaginary parts of $\phi(r)$ are shown in
Fig.~\ref{fig:wkb-exact-dipole}. The potential is very small at $r \lesssim 3a$,
which is roughly consistent with our estimate [Eq.~\eqref{eqn:R_no1}] of the no-go radius $\RnogoO(\alpha = 0.01, \beta = 3) \sim 4.6 a$.
At $r > \RnogoO$, function $\Re \mathrm{e}\, \phi(r)$ becomes approximately linear
whereas $\Im \mathrm{m}\, \phi(r)$ goes as $1 / r$, in agreement
with Eq.~\eqref{eqn:f_far}.

\begin{figure}[b]
\centering
\includegraphics[width=3.0in]{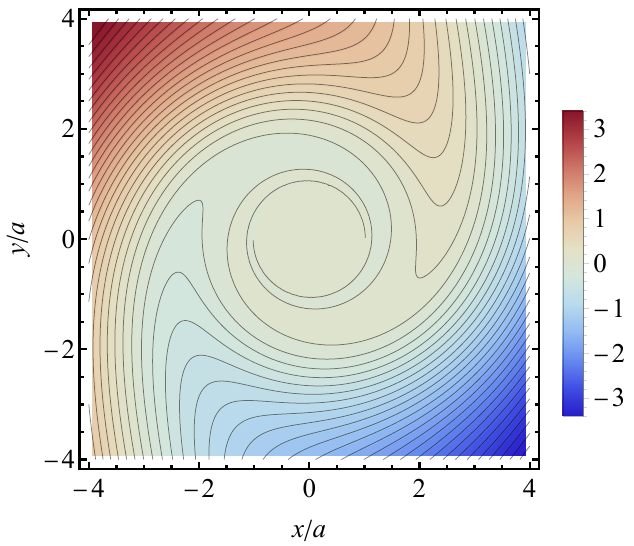}
\caption{False color and contour plots of the electrochemical potential $e\Phi(x,y)$ in the diffusive regime for the density profile given by Eq.~\eqref{eqn:denprof2}. Parameters: $\alpha = 0.01$, $\beta=3$.
\label{fig:spirals}
}
\end{figure}

A closer look at the no-go region reveals that $\phi(r)$ has some oscillations.
The period of the oscillations becomes shorter and their amplitude becomes exponentially smaller as $r$ decreases.
This fine structure is put on view more clearly in a false color plot of the 2D potential distribution $\Phi(x, y)$, Fig.~\ref{fig:spirals}.
The superimposed equipotential contours are seen to spiral inward clockwise as they approach the insulating core from $y = +\infty$ and then spiral outward counterclockwise as they go away to $y = -\infty$.
(Inside the $r < a$ core, the equipotentials go across as straight lines but this is not shown in Fig.~\ref{fig:spirals}.)
This behavior is consistent with the qualitative picture introduced in Sec.~\ref{sub:advection}.
Solutions exhibiting similar spiral patterns have also been discussed in Ref.~\cite{Fogler1995}.

In the region $r \gg a$ of primary interest we can elucidate the found exact solution by approximating it with more familiar special functions.
At such $r$ the term containing $\sigma'_{xx}$ is subleading, so we can simplify Eq.~\eqref{eqn:radial} by dropping this term. We arrive at
\begin{equation}
\phi''(r) + \frac{1}{r}\, \phi'(r)
 - \left(\frac{1}{r^2} - \frac{i\beta}{\alpha} \frac{a^{\beta}}{r^{\beta + 2}} \right) \phi(r)
 = 0 ,
\label{eqn:radial_approx}
\end{equation}
which can be interpreted as a radial Schr\"odinger equation in the $p$-wave channel for
a particle subject to an imaginary scalar potential $V(r) \propto -i r^{-\beta - 2}$.
This equation has two linearly independent solutions, one of which exponentially increases
and one decreases as $r$ approaches the origin:
\begin{equation}
\phi_\pm(r) \simeq r^{\frac\beta4} \exp \left(\pm \frac{2}{\sqrt{i \alpha \beta}} \left(\frac{a}{r}\right)^{\frac{\beta}{2}}\right),
\quad r \to 0\,.
\label{eqn:pot_near_approx}
\end{equation}
This asymptotic form can be derived using, e.g., the Wentzel-Kramers-Brillouin (WKB) approximation, see Appendix~\ref{app:WKB}.
The solution consistent with our ``no-go'' principle is the one with the minus sign
in the argument of the exponential. The no-go radius is set by $r$ such that this
argument is of the order of unity, in agreement with Eq.~\eqref{eqn:R_no1}.

The full analytical form of this approximate solution is
\begin{equation}
\phi(r) = -\frac{2 a}{\Gamma\left(\frac{2}{\beta}\right)}\,
 \left(\frac{1}{i \alpha \beta}\right)^{\frac1\beta}
K_{\frac2\beta} \left(\frac{2}{\sqrt{i\alpha \beta}} \left(\frac{a}{r}\right)^{\frac{\beta}{2}}\right),
\label{eqn:pot_besselk}
\end{equation}
where $K_\nu(x)$ is the modified Bessel function of the second kind
and $\Gamma(x)$ is the Euler Gamma function.
This expression is normalized to match Eq.~\eqref{eqn:f_far}
at large $r$.
The Landauer dipole parameters extracted from this matching according to Eq.~\eqref{eqn:R_dip} are
\begin{equation}
\frac{\Rdip^2}{a^2} =
\frac{\Gamma \left(1 - \frac{2}{\beta}\right)}
     {\Gamma \left(1 + \frac{2}{\beta}\right)}
\left(\frac{1}{\alpha \beta}\right)^{\frac2\beta} ,
\quad
\thetadip = \pi - \frac{\pi}{\beta}\,.
\label{eqn:Rdip_beta}
\end{equation}
Under the assumed condition $\beta > 2$ the size $\Rdip$ of the Landauer dipole is finite but at $\beta = 2$ it diverges.
It can be shown that at $\beta \le 2$ the far-field behavior of the potential no longer obeys Eq.~\eqref{eqn:f_far}: the correction $\phi(r) + r$ decays slower than $1 / r$.
The size of the Landauer dipole given by Eq.~\eqref{eqn:Rdip_beta} is parametrically the same as the no-go radius:
\begin{equation}
\Rdip \sim \RnogoO \sim \alpha^{-\frac1\beta} a\,.
\label{scale-ohmic}
\end{equation}
In this sense, at large distances the effect of the depletion 
is similar to that of a hard-wall circular barrier of radius $\RnogoO$.
However, there is one curious difference.
As mentioned in Sec.~\ref{sec:intro}, the Landauer dipole rotation angle
for the hard-wall case is
\begin{equation}
\thetadip = 2 \theta_H = 2\arctan\left(\frac{1}{\alpha}\right) \simeq \pi\,,
\label{eqn:R_dip_hardwall}
\end{equation}
see also Eq.~\eqref{eqn:lambda_abrupt1} below.
Comparing Eqs.~\eqref{eqn:Rdip_beta} and \eqref{eqn:R_dip_hardwall},
we discover that for the gradual density profile the angle $\thetadip$ is reduced by $\pi / \beta$ compared to that for the hard-wall barrier.

Note that the hard-wall case can be analyzed within our formalism simply by taking the limit $\beta \to \infty$ where the density profile becomes step-like: $n(r) = 0$ at $r < a$ and $n(r) = n_0$ at $r > a$.
Using the exact solutions of Eq.~\eqref{eqn:radial} rather than the small-$\alpha$ approximation of Eq.~\eqref{eqn:pot_besselk},
we can reproduce the formula $\thetadip = 2 \theta_H$ in full,
see Appendix~\ref{app:beta3}.

\subsection{Density accumulation}
\label{sub:accumulation}

Let us briefly discuss the model where the electron density is locally raised rather than depleted,
\begin{equation}
	\frac{n(r)}{n_0} \simeq
	1 + \left(\frac{a}{r}\right)^{\beta}, \quad r\gg a .
	\label{eqn:denprof3}
\end{equation}
In this case the direction of the drift velocity $\vect{v} = \hat{\vect{z}} \times \bm{\nabla}\sigma_{xy}(\vect{r})$ is reversed.
The corresponding plot of the electrochemical potential
can be obtained by reflecting Fig.~\ref{fig:spirals} across the $x$-axis.
Equations~\eqref{eqn:radial_approx}--\eqref{eqn:pot_besselk} are replaced by
their complex conjugates and the sign of $\thetadip$ in Eq.~\eqref{eqn:Rdip_beta} is reversed.

As for the hard-wall limit, the closest related model is an abrupt junction:
\begin{equation}
\begin{split}
	\sigma_{xx}(r) &= \left\{
	\begin{array}{rrr}
		\sigma_{xx}^{(1)} , & & r < a\,,
		\\
		\sigma_{xx}^{(0)} , & & r \ge a\,,
	\end{array}
	\right.
	\\
	\sigma_{xy}(r) &= \left\{
\begin{array}{rrr}
	\sigma_{xy}^{(1)} , & & r < a\,,
	\\
	\sigma_{xy}^{(0)} , & & r \ge a\,.
\end{array}
\right.
\end{split}
	\label{eqn:denprof_abrupt}
\end{equation}
Matching the outer and inner solutions, Eqs.~\eqref{eqn:f_far} and \eqref{eqn:f_near_0},
at the boundary $r = a$, we find
\begin{equation}
	\lambda = \frac{\left(\sigma_{xx}^{(0)} + i \sigma_{xy}^{(0)}\right) - \left(\sigma_{xx}^{(1)} + i \sigma_{xy}^{(1)}\right)}
	               {\left(\sigma_{xx}^{(0)} - i \sigma_{xy}^{(0)}\right) + \left(\sigma_{xx}^{(1)} + i \sigma_{xy}^{(1)}\right)}\, a^2 \,.
	\label{eqn:lambda_abrupt}
\end{equation}
If $\sigma_{xx}^{(0)} / \sigma_{xy}^{(0)} = \sigma_{xx}^{(1)} / \sigma_{xy}^{(1)} = \alpha$, then
\begin{equation}
	\lambda = -\frac{1 - i \alpha}{1 - i A \alpha}\, a^2\,,
	\quad A \equiv \frac{\sigma_{xx}^{(1)} + \sigma_{xx}^{(0)}}{\sigma_{xx}^{(1)} - \sigma_{xx}^{(0)}} \,.
	\label{eqn:lambda_abrupt1}
\end{equation}
For example, if the density accumulation is very high, $A \simeq 1$, then the Landauer dipole size $R_L \simeq a$ and angle $\thetadip \simeq -\pi$ are approximately independent of the magnetic field.
For more details, see Appendix~\ref{app:beta3}.

%=========================================================
\section{Hydrodynamic transport}
\label{sec:hydro}
%=========================================================
\subsection{Model}

We now turn to the regime where frequent electron–electron collisions endow the electron fluid with shear viscosity~\cite{Chapman1999, Steinberg1958, Kaufman1960, Alekseev2016}
\begin{equation}
 \nu = \frac14\, \frac{ v_F^2 \tau_\mathrm{ee}}{1 + 4 \omega_c^2 \tau_\mathrm{ee}^2} \,,
\label{eqn:nu}
\end{equation}
where $v_F$ is the Fermi velocity and $\tau_\mathrm{ee}$ is the average time between collisions.
(For the specific case of graphene, see Refs.~\onlinecite{Principi2016, Narozhny2017, Lucas2018, Sun2018, Kiselev2019}; for a broader theoretical approach to relativistic magneto-hydrodynamics,
see also Refs.~\cite{Amoretti2020} and \cite{Amoretti2022}.)
We focus on the range of magnetic fields such that $1 / \tau_\mathrm{mr} \ll \omega_c \ll 1 / \tau_\mathrm{ee}$ where $\nu$ is $B$-independent.
The corresponding hierarchy of the key length scales is
\begin{equation}
	l_\mathrm{ee} \ll R_c \ll l_\mathrm{mr}\,,
\end{equation}
where $l_{\rm ee} = v_F \tau_\mathrm{ee}$ is the electron–electron mean free path,
$R_c = v_F / \omega_c$ is the cyclotron radius,
and $l_{\rm mr} = v_F \tau_\mathrm{mr}$ is the momentum-relaxation length.

Equation~\eqref{eqn:E_from_j} acquires an extra viscous friction term:
\begin{equation}
\vect{E}(\vect{r}) = \rho_{xx}(\vect{r})\, \vect{j}(\vect{r})
- \rho_{xy}(\vect{r}) \left[\hat{\vect{z}} \times \vect{j}(\vect{r})\right]
- \frac{1}{n e} \, \bm{\nabla} \cdot \vect{\Pi}(\vect{r})\,,
\label{eqn:E_from_j_hydro}
\end{equation}
where $\vect{\Pi}$ is the viscous stress tensor~\cite{Lamb2005, Landau1987}.
Its components are given by
\begin{equation}
\begin{split}
    \Pi_{i k} &= m n \nu \left[\nabla_i u_k + \nabla_k u_i
    - \frac{d - 1}{d} \delta_{i k} \left(\bm{\nabla} \cdot \vect{u}\right)\right] 
    \\
    \mbox{} &+ m n \nu_b\, \delta_{i k} \left(\bm{\nabla} \cdot \vect{u}\right)
%    \\
%    \mbox{} &+ \frac{m n \nu_H}{2} \left[
%    \epsilon_{i l} (\nabla_l u_k + \nabla_k u_l)
%    + \epsilon_{k l} (\nabla_l u_i + \nabla_i u_l)
%    \right]
\,,
    \label{eqn:T_ik}
\end{split}
\end{equation}
where $d = 2$ is the space dimension, $\vect{u}$ is the flow velocity related to the current via
\begin{equation}
    \vect{j} = e n \vect{u}\,,
    \label{eqn:conthydro}
\end{equation}
and $\nu_b$ is the bulk viscosity (or ``second'' viscosity~\cite{Landau1987}).
For simplicity, we assume that $\nu_b + \frac12 \nu = 0$,
so that the total viscous friction force is just
\begin{equation}
\bm{\nabla} \cdot \vect{\Pi}(\vect{r}) = m n \nu\, \bm{\nabla}^2 \vect{u}\,.
\label{eqn:div_T}
\end{equation}
In principle, the system also possesses the Hall viscosity $\nu_H = 2 \nu \omega_c \tau_\mathrm{ee}$~\cite{Chapman1999, Steinberg1958, Kaufman1960, Alekseev2016,
Bradlyn2012, Hoyos2012, Holder2019}.
Under the stated condition $\omega_c \tau_\mathrm{ee} \ll 1$
we can neglect it.
Our Eq.~\eqref{eqn:E_from_j_hydro} and the earlier Eqs.~\eqref{eqn:j_from_E}, \eqref{eqn:E_from_j} also neglect the thermopower arising from the gradient of temperature $T$ that enters the Gibbs-Duhem relation
$\bm{\nabla} \varepsilon_F = n^{-1} \bm{\nabla} p - s \bm{\nabla} T$.
Here $s$ is the entropy per particle and $p$ is the pressure.
Thermopower may lead to experimentally observable effects
in the low-density (or $p$-$n$ junction) region $r \sim a$~\cite{Krebs2023, Krebs2026};
otherwise, it is rather small in a degenerate 2D electron fluid and it results in corrections that are of higher order in gradients
than the terms we have kept, cf.~Refs.~\cite{Andreev2011, Lucas2018, Sun2018, Kiselev2019, Xian2023}
On the other hand, Eq.~\eqref{eqn:E_from_j_hydro}
does include the pressure contribution $-p\delta_{i k}$ to the total stress tensor~\cite{Lamb2005, Landau1987}
because $\vect{E} = -\bm{\nabla} \Phi_e - \frac{1}{e}\, \bm{\nabla} \varepsilon_F$ is not the electric field
but rather the electrochemical gradient.

Requiring $\bm{\nabla} \times \vect{E} = 0$,
after some algebra, we arrive at the fourth-order partial differential equation for $\Psi$:
\begin{equation}
\left(1 - l_G^2 \bm{\nabla}^2 \right)
\bm{\nabla} \cdot \left(\rho_{xx} \bm{\nabla} \Psi\right)
+ \left(\hat{\vect{z}} \times \bm{\nabla}\rho_{xy}\right)
\cdot \bm{\nabla} \Psi = 0\,,
\label{eqn:stream_equation_hydro}
\end{equation}
where
\begin{equation}
  l_G = \sqrt{\nu \tau_{\rm mr}} \sim \sqrt{l_{\rm ee}l_{\rm mr}}
\label{eqn:l_G}
\end{equation}
is the aforementioned Gurzhi length.
As we discuss below, viscosity plays an important role if $l_G \gg a$.

\subsection{Qualitative discussion}

\begin{figure}[t]
\centering
\includegraphics[width=3.0in]{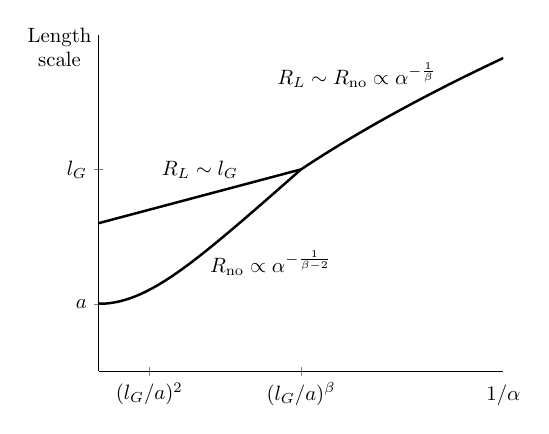}
\caption{A cartoon of the dependence of the no-go length $R_\mathrm{no}$ and the Landauer dipole
	size $\Rdip$ on $1 / \alpha$, i.e., the magnetic field.
	The crossover from the viscous regime (where $\RnogoH \ll \Rdip$) to the diffusive regime
	 (where $\RnogoO \sim \Rdip$) occurs as the field increases
	and $\RnogoH$ (or equivalently, $\Rdip$) exceeds the Gurzhi length $l_G$.
\label{fig:regimes}
}
\end{figure}

Assuming gradients scale as $\bm{\nabla}\sim 1/\RnogoH$, Eq.~\eqref{eqn:stream_equation_hydro} maps 
to the advection–diffusion equation with the diffusion coefficient and the drift velocity given by
\begin{equation}
D \sim \left(1 + \frac{l_G^2}{R^2_\mathrm{no}} \right) \rho_{xx}\,,
\quad
\vect{v} = \hat{\vect{z}} \times \bm{\nabla}\rho_{xy}\,,
\label{eqn:D_and_v}
\end{equation}
so that Eq.~\eqref{eqn:spreading1} changes to
\begin{equation}
  \delta(\RnogoH) \sim
  \sqrt{
  \left(1 + \frac{l_G^2}{R^2_\mathrm{no}} \right) \frac{\alpha a}{\beta} 
  \left(\frac{\RnogoH}{a}\right)^{\beta + 1} \RnogoH
  } \sim \RnogoH \,.
\label{eqn:spreading2}
\end{equation}
Supposing $\beta$ and $l_G \gg a$ are fixed, then as a function of $\alpha$,
the solution of Eq.~\eqref{eqn:spreading2} exhibits a change in behavior
at $1 / \alpha \sim (l_G / a)^{\beta}$.
In stronger fields, $\RnogoH$ follows Eq.~\eqref{eqn:R_no1},
with small corrections. In weaker fields, such that
\begin{equation}
\left( \frac{l_G}{a} \right)^2 \ll  \frac{1}{\alpha}
\ll \left( \frac{l_G}{a} \right)^{\beta} ,
\label{eqn:viscous_regime}
\end{equation}
the no-go length obeys a power-law with a different exponent,
$\RnogoH \propto \alpha^{-\frac{1}{\beta - 2}}$.
For future convenience, we define $\RnogoH$ with a specific numerical coefficient,
as follows:
\begin{equation}
\RnogoH = \left(\frac{\beta}{\alpha} \, \frac{a^2}{l_G^2} \right)^{\frac{1}{\beta - 2}} a
 = \left(\beta \, \frac{a^2 \omega_c}{\nu} \right)^{\frac{1}{\beta - 2}} a.
\label{eqn:R_no}
\end{equation}
In this viscosity-dominated regime,
the no-go length grows as $\RnogoH \propto B^{1/(\beta - 2)}$.
In still weaker fields, $1 / \alpha \ll (l_G / a)^2$, the no-go length should be equal to the geometric size $a$ of the depletion. This dependence is shown schematically in
Fig.~\ref{fig:regimes}.

\subsection{Far-field region}
\label{sub:Stokes}

At large distances, the resistivities $\rho_{xx}$, $\rho_{xy}$ become uniform,
so that Eq.~\eqref{eqn:stream_equation_hydro} reduces to
\begin{equation}
\bm{\nabla}^2
\left[\bm{\nabla}^2 \Psi(\vect{r}) - \frac{1}{l_G^2} \Psi(\vect{r}) \right]
 = 0\,.
\label{eqn:stokes_ns}
\end{equation}
Suppose the current is also uniform at infinity, then $\Psi(\vect{r})$ must be a
$p$-wave,
\begin{equation}
    \Psi(\vect{r}) = \Re\mathrm{e} \left[-i e^{i (\theta + \theta_H)} \psi(r)\right] ,
    \label{eqn:Psi_from_psi}
\end{equation}
with the large-$r$ behavior
\begin{equation}
\psi(r) \simeq -r + \frac{\tilde\lambda}{r}\,,
\qquad \tilde\lambda \equiv \lambda e^{-2 i \theta_H}
\label{eqn:psi_far}
\end{equation}
to match Eq.~\eqref{eqn:f_far}.
\Big[For simplicity, we omit the constant factor $\sqrt{\sigma_{xx}^2(\infty) + \sigma_{xy}^2(\infty)}\,$.\Big]
The general solution of Eqs.~\eqref{eqn:stokes_ns}--\eqref{eqn:psi_far} is
\begin{equation}
\psi(r) = - r + \frac{\tilde\lambda}{r}
+ \mu K_{1} \left({r}/{l_G}\right) ,
\label{eqn:psi_far_hydro}
\end{equation}
where $\mu$ is a constant.
In the high viscosity limit $l_G \to \infty$,
the last term behaves as $\ln r + \mathrm{ const}$, signifying the breakdown of the dipole law.
This is known as the Stokes paradox~\cite{Lamb2005, Landau1987}.
Since $K_{1} \left({r}/{l_G}\right)$ decays exponentially at $r \gg l_G$,
in the present case this paradox is moot.

The coefficients $\tilde\lambda$ and $\mu$ can be determined by matching Eq.~\eqref{eqn:psi_far_hydro} with the solution for $\psi(r)$ at short distances.
For example, if we impose the no-slip boundary condition $\psi'(r) = \psi(r) = 0$ at some $r = a$,
then~\cite{Lucas2017}
\begin{equation}
\tilde\lambda = a^2 - \mu a K_1(a/l_G) = a^2
+ 2 {a}{l_G} \frac{K_1({a}/{l_G})}{K_0({a}/{l_G})}\,.
\label{eqn:lambda_hydro_hard}
\end{equation}
Similar formulas have been derived~\cite{Gornyi2023, Kiselev2019, Alekseev2023}
for other types of boundary conditions, e.g., no-stress boundary conditions. 
The only effect of the magnetic field is to rotate the Landauer dipole by twice the Hall angle $\theta_H$ with respect to the average electric field,
\begin{equation}
\lambda(B) = \lambda(0) e^{2 i \theta_H(B)} ,
\end{equation}
same as for diffusive transport around a hard obstacle.
(If the Hall viscosity $\nu_H$ is retained,
then the flow-induced density profile is rotated by an angle of the order of $\nu_H / \nu \sim \omega_c \tau_\mathrm{ee}$~\cite{Gornyi2023}.)
If $l_G$ is much larger than $a$, we can expand the Bessel functions in Eq.~\eqref{eqn:lambda_hydro_hard} to get
\begin{equation}
 \Rdip = \left| \tilde\lambda \right|^{1/2} \simeq \frac{\sqrt{2}}{\sqrt{\ln(l_G / a)}}\, l_G\,.
 \label{eqn:Rdip_hydro_hard}
\end{equation}
This formula shows that an obstacle of radius $a$ can perturb the flow at a much larger distance $\Rdip$, which is equal to the Gurzhi length reduced by a logarithmic factor.
In our case, the no-slip boundary condition is effectively imposed at the no-go radius,
so we expect
\begin{equation}
  \Rdip^2 \simeq
  \frac{2 l_G^2}{\ln \left(l_G / \RnogoH \right)}\,,
 \quad l_G \gg \RnogoH\,.
\label{eqn:Rdip_hydro}
\end{equation}

\subsection{Analytical approximations}
\label{sub:analytical2}

Being the fourth order, the viscous hydrodynamics Eq.~\eqref{eqn:stream_equation_hydro} is much more complicated than the diffusive transport one (the same equation with $l_G = 0$).
We have not succeeded in finding its full analytical solution for either of the two studied density profiles, Eqs.~\eqref{eqn:denprof2} and \eqref{eqn:denprof1}.
An approximate solution can be derived if the three relevant length scales are well separated:
\begin{equation}
	a \ll \RnogoH \ll l_G
	\label{eqn:length_hierarchy}
\end{equation}
using the asymptotic matching method explained below.

First, as in Sec.~\ref{sub:analytical1}, we can simplify the problem by treating $\rho_{xx}$ as a constant
at distances $r \gg a$.
The equation for $\psi(r)$ becomes
\begin{equation}
\left(1 - l_G^2 \Delta_r \right)
\Delta_r \psi(r) - \frac{i \beta}{\alpha}\, \frac{a^\beta}{r^{\beta + 2}} \psi(r) = 0\,,
\label{eqn:psi_equation2}
\end{equation}
where $\Delta_r$ is the Laplacian operator in the $p$-wave channel:
\begin{equation}
\Delta_r = \frac{d^2}{d r^2} + \frac{1}{r}\,\frac{d}{d r} - \frac{1}{r^2}\,.
\label{eqn:psi_equation3}
\end{equation}
Next, at intermediate distances $a \ll r \ll l_G$, we can simplify Eq.~\eqref{eqn:psi_equation2} to
\begin{equation}
-\Delta_z^2\psi - \frac{i}{z^{\beta + 2}} \psi = 0\,,
\quad z \equiv \frac{r}{\RnogoH} \,.
\label{eqn:psi_equation4}
\end{equation}
This equation has four linearly independent solutions.
We need to select the one that obeys the correct boundary conditions.
At $z \ll 1$, i.e., $r \ll \RnogoH$,
the WKB approximation is valid. It predicts four possible types of asymptotic behavior:
\begin{equation}
\psi \sim \exp\left(
-\frac{4}{\beta - 2}\, e^{-\frac{i\pi}{8}} i^k
z^{-\frac{\beta - 2}{4}}\right) ,
\label{eqn:psi_WKB}
\end{equation}
depending on $k = 0$, $1$, $2$, or $3$. (Here $i^k$ is a quartic root of unity.)
The options $k = 0$ and $1$ describe oscillations that decay exponentially as $r
= \RnogoH z \to 0$,
so they are allowed;
$k = 2$ and $3$ correspond to an exponential increase, so they must be ruled out.

\begin{figure}[b]
	\centering
	\includegraphics[width=3.0in]{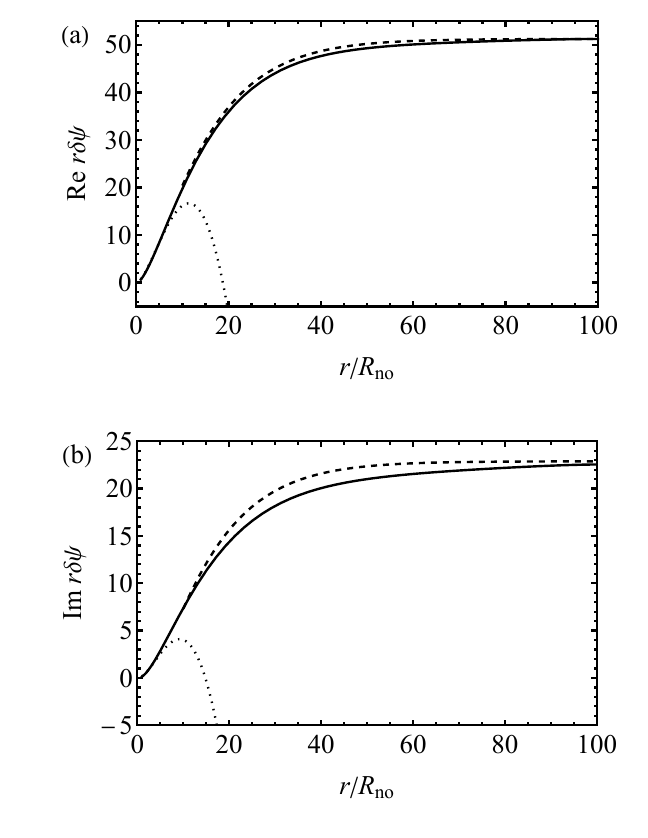}
	\caption{Analytical and numerical results for the stream function.
		The panels (a) and (b) depict, respectively, the real and imaginary parts
		of $r \delta \psi(r) \equiv r [r + \psi(r)] / \RnogoH^2$ that approaches $\tilde\lambda / \RnogoH^2 = 51.2 + 22.9i$
		at large $r$.
		The solid line is the numerical solution of Eq.~\eqref{eqn:psi_equation2}
		with the boundary conditions $\psi(r_\mathrm{min}) = \psi'(r_\mathrm{min}) = 0$,
		$\psi(r_\mathrm{max}) = -r_\mathrm{max} + \tilde\lambda / r_\mathrm{max}$,
		$\psi'(r_\mathrm{max}) = -1 - \tilde\lambda / r_\mathrm{max}^2$.
		The dotted line is obtained from Eq.~\eqref{eqn:psi_Meijer};
		the dashed line (plotted for $r > l_G$) is from Eqs.~\eqref{eqn:psi_far_hydro}, \eqref{eqn:mu_from_c}, \eqref{eqn:Rdip_hydro_beta3}. 
		Parameters: $\beta = 3$, $l_G / \RnogoH = 10$, $r_\mathrm{min}  / \RnogoH = 0.1$,
		$r_\mathrm{max}  / \RnogoH = 100$.
		\label{fig:delta_psi}
	}
\end{figure}

At distances $\RnogoH \ll r \ll l_G$, the $i / z^{\beta + 2}$ term in Eq.~\eqref{eqn:psi_equation4} can also be discarded, which yields the biharmonic equation $\Delta_z^2 \psi = 0$
with the general solution
\begin{equation}
	\psi = \frac{c_1}{z}
	 + c_2 z
	 + c_3 z \ln z
	 + c_4 z^3 .
	\label{eqn:psi_far2}
\end{equation}
The error of this approximation is about ${\psi} / {z^{\beta - 2}}$.
Equation~\eqref{eqn:psi_far2} must match Eq.~\eqref{eqn:psi_far_hydro} or more precisely, its expansion excluding terms of order $\mu /\, l_G^3$ or smaller:
\begin{equation}
\psi(r) \simeq \frac{\tilde\lambda + \mu l_G}{r} - r + \frac{\mu}{2 l_G}\, r \left(\ln \frac{r}{2 l_G} - \frac12 + \gamma\right) ,
\label{eqn:psi_intermediate}
\end{equation}
where $\gamma = 0.577\ldots$ is the Euler constant.
This yields
\begin{gather}
\mu = -\frac{2 l_G}{\frac{c_2}{c_3} + \ln \frac{2 l_G}{\RnogoH} + \frac12 - \gamma}\,,
\label{eqn:mu_from_c}
\\
c_4 = 0\,.
\label{eqn:c_4}
\end{gather}
If $\beta > 4$, we have one more matching condition $\tilde\lambda + \mu l_G = c_1 \RnogoH$.
If $\beta \leq 4$, we can only say
\begin{equation}
	\tilde\lambda + \mu l_G = O(c_2, c_3) \RnogoH
\label{eqn:c_1}
\end{equation}
based on the above error estimate at $z \sim 1$.

In Appendix~\ref{app:hydro}, we study the case $\beta = 3$ in more detail and arrive at
\begin{equation}
	\tilde\lambda
	= \frac{2 l_G^2}
	{\ln\left(\frac{2 l_G}{\RnogoH} \right) + 3 - 5 \gamma - \frac{i \pi}{2}}
	+ O\left(R_\mathrm{no}^2\right) ,
	\label{eqn:Rdip_hydro_beta3}
\end{equation}
which agrees with Eq.~\eqref{eqn:Rdip_hydro}.
From this we can also extract the Landauer dipole orientation angle:
\begin{equation}
	\thetadip = 2 \theta_H + \arg \tilde\lambda
	\simeq 2 \theta_H + \frac{\pi}{2 \ln\left(\frac{2 l_G}{\RnogoH} \right)}\,.
	\label{eqn:thetadip_hydro_beta3}
\end{equation}
Hence, $\thetadip$ is larger than $2 \theta_H$
in the viscous transport whereas the opposite is true in the diffusive one [Eq.~\eqref{eqn:Rdip_beta}].

In Fig.~\ref{fig:delta_psi} we compare these analytical predictions with
numerical results (the solid line, see the figure caption).
The quantity plotted is	the dimensionless combination
$r \delta \psi(r) \equiv r [r + \psi(r)] / \RnogoH^2$
whose asymptotic value is $\tilde\lambda / \RnogoH^2$.
The numerical results behave as expected, approaching the
dashed line [Eq.~\eqref{eqn:psi_far_hydro}] at large $r$
and the dotted line [Eq.~\eqref{eqn:psi_Meijer}] at small $r$.
All the lines merge together at $\RnogoH \lesssim r \lesssim l_G$,
validating our matching method.
Given that the parameter $l_G / \RnogoH = 10$
used in this calculation is only modestly large, the agreement can be considered good,
especially for the real part of the stream function, Fig.~\ref{fig:delta_psi}(a).
To improve the agreement for the imaginary part, we chose the second term in Eq.~\eqref{eqn:Rdip_hydro_beta3} to be
$-3 i \RnogoH^2$, which is $12\%$ of the total $\Im \mathrm{m}\, \tilde{\lambda}$.

\begin{figure}[t]
	\centering
	\includegraphics[width=3.0in]{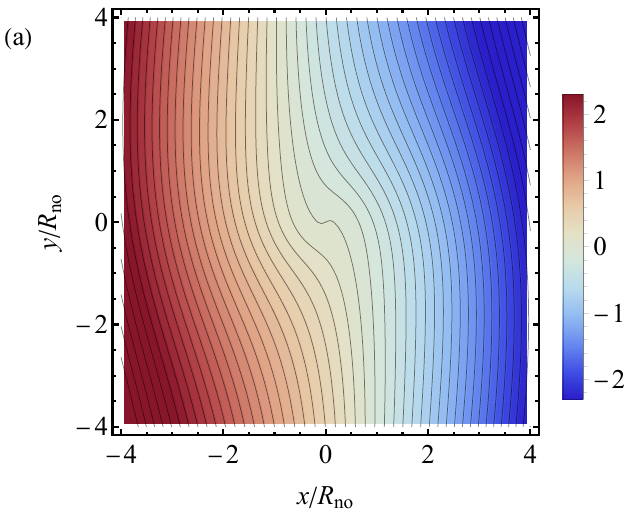}
	\includegraphics[width=3.0in]{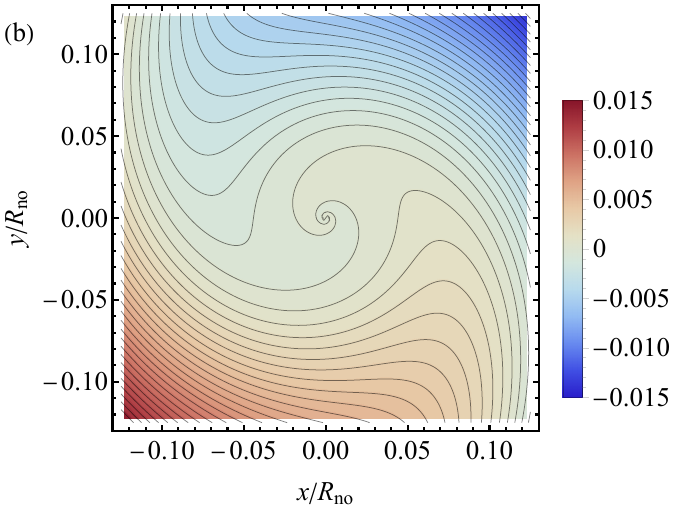}
	\caption{False color and contour plots of the stream function in the viscous regime for $\beta = 3$, $l_G/\RnogoH = 10$.
		(a)  $\Psi(x,y)$ computed using Eqs.~\eqref{eqn:psi_3_G} and \eqref{eqn:psi_Meijer}.
		(b) Enlarged view of the no-go region, computed using Eqs.~\eqref{eqn:psi_Meijer} and \eqref{eqn:psi3asym}.
		\label{fig:psi_spirals}
	}
\end{figure}

In Fig.~\ref{fig:psi_spirals} we show the 2D plots of the stream function
at intermediate distances $a \ll r \lesssim l_G$.
The zoomed-in view of the no-go region [Fig.~\ref{fig:psi_spirals}(b)] reveals spiral-like pattern of the current flow. The spiraling direction on the way from $y = +\infty$ to $y = -\infty$ is opposite to that in Fig.~\ref{fig:spirals}
because $\sin \thetadip < 0$ in the viscous case whereas $\sin \thetadip > 0$ in the diffusive case.

%=========================================================
\section{Discussion}
\label{sec:discussion}
%=========================================================

In this study, motivated by recent STP experiments~\cite{Krebs2026},
we examined a 2D magnetotransport around an inhomogeneity in the form of
a circular-symmetric depletion with a power-law electron density profile.
We confined ourselves to the case where the exponent $\beta > 2$ of this power law is sufficiently large,
so that this disturbance belongs to the class of local perturbations.
We modeled the transport by both diffusive and hydrodynamic equations.
The main qualitative result is that in strong magnetic fields
the current is repelled away from the depletion region to distances much greater than its characteristic
geometric size.
The corresponding ``no-go'' radius increases with magnetic field.
It follows the power law with a small exponent $(1 - \chi) / \beta$ in the diffusive case
and a larger exponent $1 / (\beta - 2)$ in the hydrodynamic case.
The crossover between the two scaling laws occurs when the no-go radius reaches the Gurzhi length.

The exponentially small current inside the no-go region exhibits an intriguing spiral pattern.
The spiraling is driven by the magnetic Lorentz force,
so it has a different origin from vortical flow that arises in viscous hydrodynamics~\cite{Moffatt1964, Bushong2007, Levitov2016, Bandurin2016, Falkovich2017, AharonSteinberg2022} in the absence of the magnetic field.
The twisting of the current lines at short distances has an effect on the asymptotic behavior.
It determines the angular orientation of the Landauer dipole,
i.e., the correction to the uniform current at large distances.

The Landauer dipole and the effective obstacle size we have investigated
are the simplest and the most robust characteristics.
They can be extracted already from STP maps collected at a moderate resolution.
By varying $B$ and tracking how the no-go radius grows, one can in principle distinguish diffusive and hydrodynamic behavior and estimate the electron viscosity.
Since the suppression of the current $\mathbf{j}(\mathbf{r})$ is gradual rather than abrupt,
there is some freedom in how the no-go radius $\RnogoO$ can be extracted
from numerical or experimental data. For example, it can be done in terms of a certain threshold for
the ratio $|\mathbf{j}| / |\mathbf{j}(\infty)|$.
Different thresholds would produce $\RnogoO$ varying by a numerical factor;
however, the parametric dependence of $\RnogoO$ on $B$ would stay the same.
For typical graphene devices $\RnogoO$ and $\Rdip$
are expected to range from sub-micrometer to several micrometers, depending largely
on the radius of the depletion region.

High-resolution scans~\cite{Krebs2026} have been shown to also resolve 
a small anomaly of the electrochemical potential predicted to occur~\cite{Holder2019a, Raichev2020} at the distance
of one cyclotron diameter $2 R_c$ from the hard-wall barrier.
In this regard, our classical continuum medium theory has a number of limitations.
Both the diffusive and the hydrodynamic approximations become inaccurate at short distances.
They do not account for features on the scale of the cyclotron radius where the motion is ballistic.
These approaches also fail at low temperatures and long distances,
where quantum localization --- the hallmark of the QHE --- becomes important.
(However, the latter is much less of a concern for modern high-mobility devices, where the localization length can be macroscopically large.)
To remedy these shortcomings, self-consistent solutions of the Boltzmann kinetic equation and quantum transport simulations in realistic potential profiles may be necessary.
The scaling rules we have derived may provide a useful reference
for these more ambitious calculations.

\section*{Acknowledgements}

We are grateful to Z.~Krebs and V.~Brar for prior collaboration on Ref.~\cite{Krebs2026} that inspired this study. We thank D.~G.~Polyakov for illuminating comments on the manuscript.

%=========================================================
\appendix
%=========================================================

%=========================================================
\section{Exact solutions for the diffusive transport}
\label{app:beta3}

For the density profile Eq.~\eqref{eqn:denprof2},
Eq.~\eqref{eqn:radial} can be solved in terms of the Gauss hypergeometric
function, which is a particular case of the 
generalized hypergeometric function~\cite{nist_dlmf}
 \[
 {}_p F_q\left(
 \begin{array}{c}
 	a_1, \ldots,	a_p \\
 	b_1, \ldots, b_q
 \end{array}
 \middle|\, x \right) .
 \]
 There are two linearly independent solutions, $f_\pm(r)$:
\begin{gather}
  f_\pm(r) = \left(\frac{r}{a}\right)^{\frac12 \beta c_{\pm}}
 {}_2 F_1\left(
  \begin{array}{c}
  	a_{\pm},\, b_{\pm} \\
  	 c_{\pm}
  \end{array}
  \middle|\, \frac{r^\beta}{a^\beta} \right) ,
 \label{eqn:fg_hypergeom}\\
 a_{\pm} = \tfrac{1}{2} - \tfrac{1}{\beta} \pm \kappa\,,
 \\
 b_{\pm} = \tfrac{1}{2} + \tfrac{1}{\beta} \pm \kappa\,,
 \\
 c_{\pm} = 1 \pm 2\kappa\,,
 \\
 \kappa = \frac{\sqrt{(4 + \beta^2) \alpha^2 + 4 i \alpha \beta}}{2\alpha \beta} .
\end{gather}
The solution that satisfies the proper boundary conditions is their linear combination
\begin{equation}
  \phi(r) = C_+ f_+(r) + C_- f_-(r),
  \quad r > a\,,
  \label{eqn:phi_outer}
\end{equation}
with the coefficients
\begin{equation}
  C_\pm = -\frac{\Gamma(\pm 2 \kappa)}{\Gamma\left(\tfrac{2}{\beta}\right)} \,
           \frac{\Gamma\left(b_{\mp}\right)}
                {\Gamma\left(a_{\pm}\right) }\, a\,.
\label{eqn:c_pm}
\end{equation}
Expanding $\phi(r)$ at large $r$, we find the expression for $\lambda$:
\begin{equation}
\frac{\lambda}{a^2} = 
-\frac{ \Gamma \left(1-\frac{2}{\beta }\right)}{\Gamma \left(1+\frac{2}{\beta }\right) }\,
\frac{ \Gamma \left(b_{+}\right) \Gamma \left(b_{-}\right)}
     { \Gamma \left(a_{+}\right) \Gamma \left(a_{-}\right)} \,.
\label{eqn:lambda_model2}
\end{equation}
This formula can be simplified in the two limits discussed in the main text.
For small $\alpha$ and finite $\beta$, we get
\begin{equation}
\frac{\lambda}{a^2} = 
-\frac{ \Gamma \left(1-\frac{2}{\beta }\right)}{\Gamma \left(1+\frac{2}{\beta }\right) }\,
\left(\frac{1}{i \alpha \beta}\right)^{\frac2\beta} ,
\quad \alpha \ll 1,\, \frac{1}{\beta}\,,
\label{eqn:lambda_model2_limit1}
\end{equation}
in agreement with Eq.~\eqref{eqn:Rdip_beta}.
For an arbitrary $\alpha$ and infinite $\beta$, we get
\begin{equation}
\lim_{\beta \to \infty} \lambda = -\frac{1 - i \alpha}{1 + i \alpha}\, a^2 ,
\label{eqn:lambda_model2_limit2}
\end{equation}
which matches Eq.~\eqref{eqn:R_dip_hardwall}.

For the density profile of Eq.~\eqref{eqn:denprof1}, we find
the solution for $\phi(r)$ in terms of the Kummer confluent hypergeometric function:
\begin{equation}
\begin{split}
\phi(r) &= -r\,_1 F_1\left(
\begin{array}{c}
	-\frac{1}{\beta } - \frac{i}{\alpha  \beta } \\
	1 - \frac{2}{\beta }
\end{array}
 \middle|\, \frac{a^\beta}{r^\beta} \right) 
\\
\mbox{} &- \frac{\lambda}{r}\,_1 F_1\left(
\begin{array}{c}
	\frac{1}{\beta } - \frac{i}{\alpha  \beta } \\
	1 + \frac{2}{\beta }
\end{array}
\middle|\, \frac{a^\beta}{r^\beta} \right) ,
\end{split}
\label{eqn:phi_model3}
\end{equation}
where $\lambda$ is given by
\begin{gather}
\lambda = \Lambda(\alpha, \beta) a^2 ,
\label{eqn:lambda_model3}
\\
\Lambda(\alpha, \beta) =
-\frac{ \Gamma \left(1 - \frac{2}{\beta}\right)}{\Gamma\left(1 + \frac{2}{\beta}\right) }\,
\frac{ \Gamma \left(-\frac{i}{\alpha \beta} + \frac{1}{\beta}\right)}
{ \Gamma\left(-\frac{i}{\alpha \beta} - \frac{1}{\beta}\right)} \,.
\label{eqn:Lambda}
\end{gather}
This equation agrees with Eqs.~\eqref{eqn:lambda_model2_limit1} and \eqref{eqn:lambda_model2_limit2} in the limits indicated. This confirms that the short-range details of the density profile are unimportant if the no-go radius is much larger than $a$.

\begin{figure}[b]
	\centering
	\includegraphics[width=2.5in]{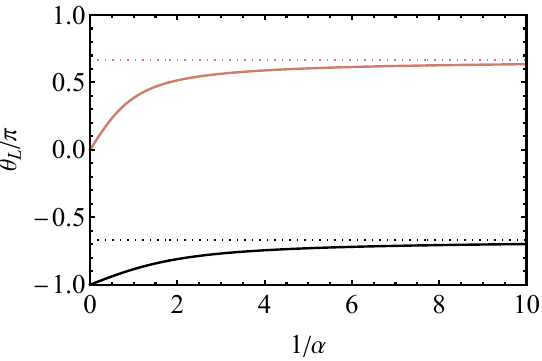}
	\caption{Landauer dipole orientation angle \textit{vs}. $1 / \alpha$ for $\beta = 3$. The red solid line is for the density depletion, Eq.~\eqref{eqn:denprof1}. The black solid line is for the density accumulation, Eq.~\eqref{eqn:denprof4}. The dashed lines show the asymptotic values $\pm \pi (1 - 1 / \beta)$.
	\label{fig:lambda}
	}
\end{figure}

The exact solution of Eq.~\eqref{eqn:radial} can be also obtained for the density profile
\begin{equation}
	n(r) = n_0 \exp \left[ \left({a}/{r}\right)^{\beta} \right] ,
	\label{eqn:denprof4}
\end{equation}
which models a density accumulation (Sec.~\ref{sub:accumulation}).
Although $n(r)$ goes to infinity at the origin,
it does not matter because this divergence is located deep inside the no-go region.
Indeed, a physically acceptable solution for $\phi(r)$ is given by Eq.~\eqref{eqn:phi_model3}
with $a^\beta$ replaced by $-a^\beta$.
This potential and the corresponding electric current are
exponentially small inside the no-go region and are regular at $r = 0$ if
Eq.~\eqref{eqn:lambda_model3} is replaced by
\begin{equation}
	\lambda \to
	\frac{1 - i \alpha}{1 + i \alpha}\,\lambda^* 
	= \frac{1 - i \alpha}{1 + i \alpha}\, \Lambda(-\alpha, \beta) a^2 ,
	\label{eqn:lambda_model3_accum}
\end{equation}
where the star denotes the complex conjugation.
If the magnetic field is strong enough, $1 / \alpha \gg \beta$,
then $\thetadip \simeq -\pi + \pi / \beta$ for the density accumulation case,
as mentioned in Sec.~\ref{sub:accumulation}.
To illustrate its dependence on the magnetic field, we plot $\thetadip$ as a function of $1 / \alpha$
in Fig.~\ref{fig:lambda} for both density depletion and accumulation cases.

%=========================================================
\section{WKB approximation}
\label{app:WKB}

To solve Eq.~\eqref{eqn:radial} for an arbitrary conductivity profile $\sigma_{x x}(r)$, $\sigma_{x y}(r)$
we can use the WKB method.
Following the procedure outlined in Sec.~5 of Ref.~\cite{Berry1972},
we seek the solution in the form of a linear combination
\begin{equation}
	\phi(r) =  C_+ f_+^\mathrm{WKB}(r) + C_- f_-^\mathrm{WKB}(r)
	\label{eqn:phi_WKB_far}
\end{equation}
of two basis functions
\begin{equation}
f_{\pm}^\mathrm{WKB}(r) = \frac{1}{\sqrt{r \sigma_{xx}(r) p(r)}}
 e^{\pm i w(r)} ,
\label{eqn:WKB_diffusive1}
\end{equation}
where
\begin{equation}
	w(r) = \int\limits^r_{r_0}  p(\rho) d\rho
	%	\simeq \int\limits^r_{r_0}
	%	\sqrt{\frac{i}{\rho}\, \frac{\sigma'_{x y}(\rho)}{\sigma_{x x}(\rho)} - \frac{1}{\rho^2}}\,
	%	{d \rho}
	\label{eqn:w}
\end{equation}
is the ``phase'' and $p(r)$ defined by
\begin{equation}
	\begin{split}
		p^2(r) &= \frac{i}{r} \frac{\sigma'_{xy}}{\sigma_{xx}}
		- \frac{1}{2} \left[ \frac{1}{r} \frac{\sigma'_{xx}}{\sigma_{xx}} + \frac{\sigma''_{xx}}{\sigma_{xx}}
		- \frac{1}{2} \left( \frac{\sigma'_{xx}}{\sigma_{xx}} \right)^2 \right]
		\\
		\mbox{} &- \frac{1}{r^2}
	\end{split}
	\label{eqn:WKB_diffusive2}
\end{equation}
is the ``momentum.''
The lower integration limit $r_0$ in Eq.~\eqref{eqn:w} can be chosen arbitrarily as long as the integral converges.
Different choices of $r_0$ can be absorbed into the coefficients $C_\pm$.
The WKB approximation amounts to treating $C_\pm$ as constants.

At large positive $r$, we find $p = \sqrt{p^2} \simeq i / r$ and $w(r) \simeq i \ln r + \text{const}$.
The corresponding $\phi(r)$ has the correct asymptotic behavior [Eq.~\eqref{eqn:f_far}]
with $\lambda$ related to the ratio of $C_+$ and $C_-$:
\begin{equation}
	\lambda = \frac{C_+ }{C_-}\, r_0^2 
	\exp \left\{2i \int\limits^\infty_{r_0}
	\left[p(\rho) - \frac{i}{\rho}\right] d\rho \right\}
	\,.
\label{eqn:lambda_WKB}
\end{equation}
The WKB approximation is also correct
to the leading order in parameter $\alpha = \sigma_{x x} / \sigma_{x y} \ll 1$ at short distances
where $\frac{d}{d r}\, |p^{-1}(r)| \ll 1$.
However, it errs at intermediate $r$ such that
\begin{equation}
	\frac{d}{d r}\, |p^{-1}(r)| \sim \frac{1}{r}\, |p^{-1}(r)| \sim
	\sqrt{\frac{1}{r}\, \frac{\sigma_{x x}(r)}{\sigma'_{x y}(r)}} \sim 1\,.
	\label{eqn:WKB_diffusive_validity}
\end{equation}
This is the same as Eq.~\eqref{eqn:difflaw} for $\RnogoO$.
Accordingly, the WKB approximation is valid if $r \ll \RnogoO$ or $r \gg \RnogoO$.
In the former region,
only function $f_-^\mathrm{WKB}(r)$ is physically acceptable.
Function $f_+^\mathrm{WKB}(r)$ grows exponentially as $r$ decreases,
so it must be discarded, i.e., $C_+ = 0$ at such $r$.
Additionally, the first term in Eq.~\eqref{eqn:WKB_diffusive2}
dominates, so that the WKB solution can be simplified to
\begin{equation}
\begin{split}
	&\phi(r) \propto \left(\frac{1}{r \sigma_{x x} \sigma'_{x y}} \right)^{\frac14}
	\exp\left[\frac{1}{\sqrt{i}} \int\limits^r_{r_0}
	\sqrt{\frac{1}{\rho}\, \frac{\sigma'_{x y}(\rho)}{\sigma_{x x}(\rho)}}\,
	d \rho \right]
	\\
	&\text{if } r \ll \RnogoO\,.
\end{split}
	\label{eqn:WKB_diffusive}
\end{equation}
This formula provides a simple analytical approximation to the exact solution
for a broad variety of conductivity profiles $\sigma_{x x}(r)$, $\sigma_{x y}(r)$.
For example, it yields Eq.~\eqref{eqn:pot_near_approx} of the main text.

By saying that the WKB approximation errs at intermediate distances
we mean that the coefficients $C_\pm$ vary significantly across this domain.
Indeed, while $C_+$ is zero at $r \ll \RnogoO$, it must be nonzero at $r \gg \RnogoO$
if $\lambda \neq 0$, cf.~Eq.~\eqref{eqn:lambda_WKB}.
The variation of $C_+$ as a function of $r$
can be studied by the method of complex trajectories, which involves
examining the behavior of $p(r)$ and $w(r)$
in the complex plane of $r$~\cite{Berry1972}.
Unfortunately, it is not possible to derive, e.g.,
Eq.~\eqref{eqn:Rdip_beta} by this method
due to the lack of so-called good paths 
where the condition
$\frac{d}{d r}\, |p^{-1}(r)| \ll 1$ is valid throughout.
Remarkably, for this specific example, we can still obtain
the second part of Eq.~\eqref{eqn:Rdip_beta},
\begin{equation}
	\arg \lambda = \pi - \frac{\pi}{\beta}\,,
	\label{eqn:thetadip_beta}
\end{equation}
by using a certain ``bad'' path.
Indeed, the equation to solve is Eq.~\eqref{eqn:radial_approx}.
Consider the complex ray $r = |r| e^{-i \delta}$, $\delta = \pi / (2 \beta)$.
On this ray, Eq.~\eqref{eqn:radial_approx} transforms to
a differential equation with real coefficients,
which can be interpreted as a radial Schr\"odinger equation with a repulsive potential $V \propto |r|^{-\beta - 2}$.
The required solution can be restricted to the real domain.
Such a solution, at large $|r|$, must behave as
\begin{equation}
	\phi(r) \propto -|r| + \frac{|\lambda|}{|r|}
	 \propto -r + \frac{|\lambda| e^{-2 i \delta}}{r}\,.
	\label{eqn:phi_complex_r}
\end{equation}
Comparing with Eq.~\eqref{eqn:f_far}, we
see that $\lambda = -|\lambda| e^{-2 i \delta}$, which is the same as Eq.~\eqref{eqn:thetadip_beta}.

Note that Eq.~\eqref{eqn:radial_approx} also transforms into a real differential equation on the ray $r = |r| e^{i \delta}$
where the effective potential $V \propto -|r|^{-\beta - 2}$ is attractive.
On this other ray, $f_-^\mathrm{WKB}(r)$ and therefore $\phi(r)$ are complex-valued oscillating functions. We cannot impose the reality condition on $\phi(r)$,
which explains why the first equation in Eq.~\eqref{eqn:phi_complex_r} is invalid
on this alternative ``bad'' path.

The WKB method can be applied to Eq.~\eqref{eqn:stream_equation_hydro} as well.
Since this equation is of the forth order,
the WKB basis consists of four functions:
\begin{equation}
	f_{k}^\mathrm{WKB}(r) = \frac{1}{\sqrt{r \rho_{xx}(r) p^3(r)}}
	\exp \left[i^k w(r)\right] ,
	\label{eqn:WKB_viscous}
\end{equation}
where $k = 1$, $2$, $3$, or $4$, and $w(r)$ is given by Eq.~\eqref{eqn:w}.
At $r \ll \RnogoH$, function $p(r)$ can be approximated by
\begin{equation}
		p(r) \simeq e^{-\frac{i\pi}{8}} \left[\frac{1}{l_G^2 r}\, \frac{\rho'_{x y}(r)}{\rho_{x x}(r)}\right]^{\frac14} .
	\label{eqn:p_viscous}
\end{equation}
For the example studied in Sec.~\ref{sub:analytical2},
this gives Eq.~\eqref{eqn:psi_WKB}.
As in the diffusive case,
the WKB method alone cannot bridge short- and long-distance behaviors,
Eqs.~\eqref{eqn:WKB_viscous} and \eqref{eqn:psi_far}, respectively.

%=========================================================
\section{Solutions for the viscous hydrodynamic transport}
\label{app:hydro}
%=========================================================

The four linearly independent solutions of Eq.~\eqref{eqn:psi_equation4} with $\beta = 3$ are
\begin{align}
\psi_1(r) &= \frac{1}{z}\, {}_0 F_3\left(
\begin{array}{c}
	- \\
	3, 3, 5
\end{array}
\middle|\, -\frac{i}{z}\right) ,
\\
\psi_2(r) &= G_{0,4}^{2,0}\left(
\begin{array}{c}
	- \\
	-1,1; -3,-1
\end{array}
\middle|\, -\frac{i}{z} \right) ,
\\
\psi_3(r) &= G_{0,4}^{3,0}\left(
\begin{array}{c}
	- \\
	-1,-1, 1;-3
\end{array}
\middle|\, +\frac{i}{z} \right) ,
\label{eqn:psi_3_G}
\\
\psi_4(r) &= G_{0,4}^{4,0}\left(
\begin{array}{c}
	- \\
	-3,-1,-1, 1; 
\end{array}
\middle|\, -\frac{i}{z} \right),
\label{eqn:psi_beta3}
\end{align}
where
\[
G_{p q}^{m n}\left(
\begin{array}{c}
a_1, \ldots,	a_p \\
b_1, \ldots, b_q
\end{array}
\middle|\, x \right)
\]
is the Meijer $G$-function~\cite{nist_dlmf} and
\begin{equation}
z = \frac{\alpha}{3}\, \frac{l_G^2}{a^3}\, r = \frac{r}{\RnogoH}\,.
\label{eqn:z_beta3}
\end{equation}
The small-$r$ asymptotic behavior of these functions is given by
Eq.~\eqref{eqn:psi_WKB}:
\begin{equation}
	\psi(r) \sim \exp\left[-4 e^{-\frac{i\pi}{8}}
	i^k \left(\frac{\RnogoH}{r} \right)^{\frac{1}{4}}\right] .
	\label{eqn:psi_WKB_beta3}
\end{equation}
Functions $\psi_1(r)$ and $\psi_2(r)$ are exponentially increasing as $r \to 0$,
so they are not allowed.
Functions $\psi_3(r)$ and $\psi_4(r)$ correspond, respectively, to $k = 1$ and $0$, i.e.,
the slower and the faster exponential decay. We need to examine them further.
At large $r$, we find
\begin{align}
	\psi_3(r) &\simeq -\frac{i}{2} \left(\ln z + \frac{5}{2} - 4 \gamma - \frac{i \pi}{2}\right) z,
	\label{eqn:psi_3_far}
\\
	\psi_4(r) &\simeq -6 i z^3 .
	\label{eqn:psi_4_far}
\end{align}
Therefore, we must discard $\psi_4(r)$ and keep $\psi_3(r)$ only.
The properly normalized expression is
\begin{equation}
\psi(r) = i \frac{\RnogoH}{l_G}\, \mu \psi_3(z)
\label{eqn:psi_Meijer}
\end{equation}
and the ratio ${c_2} / {c_3}$ needed for Eq.~\eqref{eqn:Rdip_hydro_beta3} is
\begin{equation}
	\frac{c_2}{c_3} = \frac{5}{2} - 4 \gamma - \frac{i \pi}{2}\,.
\end{equation}
The higher-accuracy asymptotic formula for $\psi_3(r)$
used in making Fig.~\ref{fig:psi_spirals}(b) is~\cite{nist_dlmf}
\begin{equation}
\begin{split}
\psi_3(z \RnogoH) &=	
e^{-\frac{9 i \pi }{16}} \sqrt{\frac{\pi}{2}}\,
\left(1 + \frac{123}{32} e^{-\frac{5 i \pi }{8}} z^{\frac{1}{4}}\right)
\\
\mbox{} &\times z^{11 / 8}
\exp\left(-4 e^{\frac{3 i\pi}{8}}
z^{-\frac{1}{4}}\right) .
\end{split}
\label{eqn:psi3asym}
\end{equation}
The exponent $11/8$ is consistent with Eq.~\eqref{eqn:WKB_viscous}.

%
%\bibliography{NOGO}

\end{document}